\documentstyle[12pt]{article}

\newcommand{\be}{\begin{equation}}
\newcommand{\ee}{\end{equation}}

\newcommand{\br}{{\bf r}}
\newcommand{\bk}{{\bf k}}

\newcommand{\vp}{\varphi}
\newcommand{\ep}{\varepsilon}
\newcommand{\al}{\alpha}
\newcommand{\ra}{\rightarrow}
\newcommand{\sgm}{\sigma}

\newcommand{\om}{\omega}

\newcommand{\Gm}{\Gamma}

\newcommand{\rgl}{\rangle}
\newcommand{\lgl}{\langle}

\begin{document}

\begin{center}
{\Large{\bf Fluctuation indices for atomic systems with
Bose-Einstein condensate } \\ [5mm]

V.I. Yukalov} \\ [5mm]

{\it Bogolubov Laboratory of Theoretical Physics, \\
Joint Institute for Nuclear Research, Dubna 141980, Russia \\
and \\
National Institute of Optics and Photonics, \\
University of S\~{a}o Paulo, S\~{a}o Carlos 13560-970, Brazil}
\end{center}

\vskip 3cm

\begin{abstract}
The notion of fluctuation indices, characterizing thermodynamic
stability of statistical systems, is advanced. These indices are
especially useful for investigating the stability of nonuniform
and trapped atomic assemblies. The fluctuation indices are
calculated for several systems with Bose-Einstein condensate.
It is shown that: the ideal uniform Bose-condensed gas is
thermodynamically unstable; trapped ideal gases are stable for
the confining dimension larger than two; trapped gases, under
the confining dimension two, are weakly unstable; harmonically
trapped gas is stable only for the spatial dimension three;
one-dimensional harmonically trapped gas is unstable;
two-dimensional gas in a harmonic trap represents a marginal case,
being weakly unstable; interacting nonuniform three-dimensional
Bose-condensed gas is stable. There are no thermodynamically
anomalous particle fluctuations in stable Bose-condensed systems.
\end{abstract}

\vskip 3cm


\vskip 3cm


\newpage

\section{Introduction}

Thermodynamic stability of statistical systems is an important
notion characterizing the possibility of existence of equilibrium
systems as such. There are several stability conditions that
are required to be fulfilled in order that the system be
thermodynamically stable [1]. In the present paper, we concentrate
on the stability related to fluctuations of observable quantities.
These fluctuations are to be {\it thermodynamically normal} as soon
as one assumes that the considered system is in thermal equilibrium.
In the other case, if the fluctuations of at least one of the
observables are thermodynamically anomalous, this implies that
this observable cannot be measured and the system equilibrium
is actually destroyed by such fluctuations, which explains their
naming as anomalous.

The problem of instability, caused by fluctuations, has recently
attracted great attention with respect to nonuniform confined
systems, such as trapped atomic gases. Especially intensive
discussions on the type of particle fluctuations have accompanied
the study of systems with Bose-Einstein condensate. Description of
the main properties of the latter systems can be found in the book
[2] and review articles [3-10].

In discussions of particle fluctuations in Bose-condensed systems,
there have been the widely spread heresy that these fluctuations
could be anomalous, been drastically different from such fluctuations
in other systems. This controversy has been investigated in detail
in Refs. [4,10-13], where it has been explained that the appearance
of such anomalous fluctuations is merely due to calculational errors.

When analyzing whether fluctuations of observable quantities are
thermodynamically normal or anomalous, it is necessary to resort
to the notion of thermodynamic limit. In the present paper, a novel
characteristic is introduced, allowing for a convenient {\it
quantitative} description of the type of the limiting behaviour
of fluctuations under the increasing number of particles. This
is the {\it fluctuation index} associated with the considered
observable. The value of this index defines when the fluctuations
are thermodynamically normal and when are not, thus, showing whether
the system is thermodynamically stable or unstable. The notion is
illustrated by calculating the fluctuation indices for different
Bose-condensed systems: confined in a box, trapped by power-law
potentials, and generally nonuniform.

\section{Fluctuation indices}

\subsection{Definition and stability conditions}

The standard definition of thermodynamic limit implies that the
number $N$ of particles in the system and its volume $V$ tend to
infinity so that
\be
\label{1}
 N \ra \infty \; , \qquad V \ra \infty \; , \qquad
\frac{N}{V} \ra const \; .
\ee
This definition assumes that there is a well defined volume of
the system.

However, for systems confined in trapping potentials, the volume
may be not fixed. In such a case, it is necessary to resort to a
more general definition of thermodynamic limit. The latter can be
done as follows [10,14]. Let $A_N$ be an extensive observable
quantity for a system with the number of particles $N$. Then the
thermodynamic limit, related to this observable, is defined as
\be
\label{2}
N \ra \infty \; , \qquad A_N \ra \infty \; ,
\qquad \frac{A_N}{N} \ra const \;  .
\ee
From here, it is seen that
\be
\label{3}
 \lim_{N\ra\infty} \; \frac{\ln|A_N|}{\ln N} \leq 1 \;  .
\ee
In the particular case, when the system volume $V$ is well defined,
hence the extensive observable $A_N \propto V$, definition (2)
reduces to the standard form (1).

Fluctuations, associated with an observable, represented by a
self-adjoint operator $\hat{A}$, are described by the dispersion,
or variance, of this operator, given by the statistical average
$$
 {\rm var} ( \hat A) \equiv 
\lgl (\hat A - \lgl \hat A \rgl )^2\rgl
= \lgl \hat A^2 \rgl - \lgl \hat A\rgl^2 \;  .
$$
The {\it fluctuation index}, related to the observable quantity
$\langle \hat{A} \rangle$, represented by a self-adjoint operator
$\hat{A}$, is defined as the limit
\be
\label{4}
\vp(\hat A) \equiv \lim_{N\ra\infty} \;
\frac{\ln{\rm var}(\hat A)}{\ln N} \;  .
\ee
Since the variance itself is an extensive observable, it has to
satisfy the {\it stability conditions}
\be
\label{5}
\vp(\hat A) \leq 1 \; , \qquad
\lim_{N\ra\infty} \; \frac{{\rm var}(\hat A)}{ N} <
\infty \;   ,
\ee
in agreement with Eq. (3). The first of these conditions is
necessary, characterizing the behavior of fluctuations in
thermodynamic limit, while the second condition is necessary and
sufficient. When both these conditions are valid, the system is
stable. It may happen that the first condition is valid, but the
second is not. In the latter case, we shall say that the system
is {\it weakly unstable}.

Fluctuations, satisfying Eq. (5), are called {\it thermodynamically
normal}, while fluctuations, for which conditions (5) are not valid,
are termed {\it thermodynamically anomalous}. This is because an
observable with normal fluctuations can be measured, but that with
anomalous ones, cannot, as soon as the magnitude of fluctuations,
hence the measurement uncertainty, is larger than the observable
itself. A statistical system, where at least one of the observables
exhibits anomalous fluctuations, cannot be in thermal equilibrium.
Such a system is thermodynamically unstable.

\subsection{Uniform ideal gas}

Let us consider particle fluctuations in a system with Bose-Einstein
condensate. The particle-number operator $\hat{N}$ is the sum
$$
\hat N = \hat N_0 + \hat N_1
$$
of the terms corresponding to condensed particles ($\hat{N}_0$)
and uncondensed particles ($\hat{N}_1$). In thermodynamic limit,
by the Bogolubov theorem [15], the operator $\hat{N}_0$ becomes
a nonoperator number, so that
$$
 {\rm var} (\hat N_0 ) \ra 0 \qquad (N \ra \infty)  \; .
$$
This fact is discussed in detail in review [8]. Therefore, particle
fluctuations are completely due to uncondensed particles,
\be
\label{6}
{\rm var} (\hat N)  =  {\rm var} (\hat N_1) .
\ee

For the ideal uniform Bose-condensed gas, confined in a volume $V$,
the condensation temperature is
\be
\label{7}
 T_c = \frac{2\pi}{m} \left [ \frac{\rho}{\zeta(d/2)}
\right ]^{2/d}  \; ,
\ee
where $d$ is space dimensionality, $\rho$ is average density, and
$\zeta(\cdot)$ is the Riemann zeta function. Below this temperature,
the condensate fraction behaves as
\be
\label{8}
n_0 = 1 - \left ( \frac{T}{T_c} \right )^{d/2} \qquad
(T \leq T_c ) \;  .
\ee
However, a finite critical temperature (7) exists only for
$d\geq 3$. This is a general feature of the absence of spontaneous
breaking of continuous symmetry in low-dimensional systems [16].

In the three-dimensional space, one has
\be
\label{9}
 {\rm var} (\hat N_1) = \left ( \frac{mT}{\pi} \right )^2
V^{4/3} \;  .
\ee
Taking into account Eq. (6) gives the fluctuation index
\be
\label{10}
\vp(\hat N) = \frac{4}{3}
\ee
that contradicts the stability condition (5). This tells us
that the ideal uniform Bose-condensed gas is unstable, being a
pathological object with thermodynamically anomalous particle
fluctuations.

\subsection{Trapped Bose gas}

An interesting question is whether the ideal Bose-condensed gas
could be stabilized being trapped in an external potential. The
most often used shape of the trapping potential is of the power-law
form
\be
\label{11}
 U(\br) = \sum_{\al=1}^d \frac{\om_\al}{2} \left |
\frac{r_\al}{l_\al} \right |^{n_\al} \; ,
\ee
in which $n_\alpha > 0$ and the trap frequencies and characteristic
lengths are connected by the relations
\be
\label{12}
\om_\al = \frac{1}{m l_\al^2} \; , \qquad
l_\al = \frac{1}{\sqrt{m\om_\al} } \;   .
\ee
It is convenient to introduce the effective trap frequency and
length by the geometric averages
\be
\label{13}
 \om_0 \equiv \left ( \prod_{\al=1}^d \om_\al\right )^{1/d} =
\frac{1}{ml_0^2} \; , \qquad
l_0 \equiv \left ( \prod_{\al=1}^d l_\al\right )^{1/d} =
\frac{1}{\sqrt{ m\om_0} } \;  .
\ee

Another important quantity, defining the confining power of
potential (11), is the {\it confining dimension}
\be
\label{14}
s \equiv \frac{d}{2} + \sum_{\al=1}^d \frac{1}{n_\al} \;   ,
\ee
where $d$ is the real-space dimension.

The properties of the ideal trapped Bose gas can be accurately
described by the generalized quasiclassical approximation [14].
Bose-Einstein condensation occurs at the critical temperature
\be
\label{15}
 T_c = \left [ \frac{N}{Bg_s(1)} \right ]^{1/s} \; ,
\ee
in which
\be
\label{16}
 B \equiv \frac{2^s}{\pi^{d/2}} \prod_{\al=1}^d
\frac{\Gm(1+1/n_\al)}{\om_\al^{1/2+1/n_\al} } \;  ,
\ee
and the generalized Bose function
\be
\label{17}
 g_s(z) \equiv \frac{1}{\Gm(s)} \int_{u_0}^\infty
\frac{zu^{s-1}}{e^u-z} \; du
\ee
is introduced [14], with the lower limit in the integral being
\be
\label{18}
 u_0 \equiv \frac{\om_0}{2T} \; .
\ee
Note that the standard Bose function corresponds to the limiting
case of $u_0 = 0$. Below the critical temperature (15), the
condensate fraction is
\be
\label{19}
n_0 = 1 - \left ( \frac{T}{T_c} \right )^s \qquad
( T \leq T_c) \;  .
\ee

But the formal occurrence of a critical temperature does not
necessarily mean the real existence of a stable Bose-condensed
system. To check the stability with respect to particle fluctuations,
we have to calculate the related fluctuation index. The trapping
potential (11) extends to infinity, so that the system volume is
not fixed. Hence, the general form of thermodynamic limit (2) is
to be employed. As an extensive quantity, we can take the internal
energy $E_N$, considering the thermodynamic limit in the form
\be
\label{20}
 N \ra \infty \; , \qquad E_N \ra \infty \; , \qquad
\frac{E_N}{N} \ra const \;  .
\ee
For the internal energy, we find
\be
\label{21}
 E_N = Bs g_{1+s}(1) T^{1+s} \; ,
\ee
which transforms limit (20) into
\be
\label{22}
 N \ra \infty \; , \qquad B \ra \infty \; , \qquad
\frac{B}{N} \ra const \; .
\ee

For the usual case of unipower trapping potentials, when $n_\al=n$,
the confining power (14) is
\be
\label{23}
s = \left ( \frac{1}{2} + \frac{1}{n} \right ) d \; .
\ee
And quantity (16) becomes
\be
\label{24}
 B = \frac{2^s}{\pi^{d/2} \om_0^s} \;
\Gm^d\left ( 1 + \frac{1}{n} \right ) \; .
\ee
Then the thermodynamic limit (22) reduces to
\be
\label{25}
 N \ra \infty \; , \qquad \om_0 \ra 0 \; , \qquad
N\om_0^s \ra const \; .
\ee

Looking at limit (22), there arises a temptation to treat the
quantity $B$ as an effective volume. The latter, however, is not
uniquely defined. And the most important is that such a quantity
$B$ cannot be used as a thermodynamic variable. Attempting to use
it as such would lead to inconsistent thermodynamic relations. That
is, though $B$ reminds something like an effective volume, there is
no any sense of identifying it with the latter.

Under the thermodynamic limit (22), or (25), the critical
temperature (15) behaves as
$$
T_c \propto N^{1-1/s} \ra 0 \qquad ( s< 1 ) \; ,
$$
$$
T_c \propto (\ln N)^{-1} \ra 0 \qquad ( s = 1 ) \; ,
$$
\be
\label{26}
 T_c \ra const \qquad (s > 1 ) \; .
\ee
For the confining dimension $s \leq 1$ the critical temperature
tends to zero. Hence, only $s > 1$ provides a finite critical
temperature.

Again, the occurrence of a condensation temperature does not
guarantee the existence of a stable Bose-condensed system. We need
to find the fluctuation indices. For the variance of the particle
number, we get
\be
\label{27}
 {\rm var} (\hat N) = \frac{g_{s-1}(1)}{g_s(1)}
\left ( \frac{T}{T_c} \right )^s N \; .
\ee
This becomes negative for $s < 1$, which contradicts the definition
of the variance as a non-negative quantity. Thus, only $s \geq 1$
can be considered. With the generalized Bose function (17), we find
$$
 {\rm var} (\hat N) = 2 \left ( \frac{T}{\om_0} \right )^2
\qquad ( s = 1) \; ,
$$
$$
 {\rm var} (\hat N) = \frac{N}{(2-s)\zeta(s)\Gm(s-1)}
\left ( \frac{2T_c}{\om_0} \right )^{2-s}
\left ( \frac{T}{T_c}\right )^2
\qquad ( 1 < s < 2) \; ,
$$
$$
 {\rm var} (\hat N) = \frac{N}{\zeta(2)}
\left ( \frac{T}{T_c}\right )^2
\ln \left ( \frac{2T_c}{\om_0} \right ) \qquad ( s = 2) \; ,
$$
\be
\label{28}
{\rm var} (\hat N) = \frac{\zeta(s-1)}{\zeta(s)}
\left ( \frac{T}{T_c}\right )^s N   \qquad ( s > 2) \; .
\ee
This yields the fluctuation indices
$$
\vp(\hat N) = 2 \qquad ( s = 1 ) \; ,
$$
$$
\vp(\hat N) = \frac{2}{s} \qquad ( 1 < s < 2 ) \; ,
$$
$$
\vp(\hat N) = 1 + 0 \qquad ( s = 2 ) \; ,
$$
\be
\label{29}
 \vp(\hat N) = 1 \qquad ( s > 2 ) \;  ,
\ee
where the notation
$$
\lim_{N\ra\infty} \; \frac{\ln \ln N}{\ln N}
\equiv + 0
$$
is used.

Consequently, ideal trapped gas can form an absolutely stable
Bose-condensed system only for $s > 2$. The case $s = 2$ is on the
boundary of stability. Strictly speaking, ${\rm var}(\hat{N})$ diverges
as $N$ tends to infinity, but this divergence is weak, being of
logarithmic type. This means that the Bose-condensed system, with
the confining dimension $s = 2$, is weakly unstable.

Remembering definition (14) gives the stability condition
\be
\label{30}
 \frac{d}{2} + \sum_{\al=1}^d \frac{1}{n_\al} > 2 \;  .
\ee

\subsection{Harmonic trapping potential}

The most commonly considered shape of trapping potentials is that of
harmonic potential, when $n_\alpha = 2$. Then $s=d$ and $B =1/\om^d$.
The condensation temperature (15) gives
$$
T_c = \frac{N\om_0}{\ln(2N)} \qquad ( d = 1) \; ,
$$
\be
\label{31}
T_c = \om_0 \left [ \frac{N}{\zeta(d)}\right ]^{1/d} \qquad
( d \geq 2 ) \; .
\ee

One often states that the formal existence of the critical
temperature $T_c$ implies the possibility of getting Bose-Einstein
condensate in one- and two-dimensional harmonic traps. But, as has
been stressed above, the mere occurrence of $T_c$ does not guarantee
that such a Bose-condensed system would be stable, hence, could really
exist. We have to check the system stability.

The thermodynamic limit (25), for harmonic traps, takes the form
\be
\label{32}
 N \ra \infty \; , \qquad \om_0 \ra 0 \; , \qquad
N\om_0^d \ra const \; .
\ee
Variance (27) yields
$$
{\rm var}(\hat N) = 2 \left ( \frac{T}{\om_0} \right )^2 \qquad
(d = 1) \; ,
$$
$$
{\rm var}(\hat N) = \left ( \frac{T}{\om_0} \right )^2
\ln \left ( \frac{2T}{\om_0} \right ) \qquad (d = 2) \; ,
$$
\be
\label{33}
{\rm var}(\hat N) = \frac{\pi^2 N}{6\zeta(3)}
 \left ( \frac{T}{T_c} \right )^3  \qquad (d = 3) \; .
\ee
As a result, the fluctuation indices are
$$
\vp(\hat N) = 2 \qquad ( d = 1 ) \; ,
$$
$$
\vp(\hat N) = 1 + 0 \qquad ( d = 2 ) \; ,
$$
\be
\label{34}
\vp(\hat N) = 1 \qquad ( d = 3 ) \; ,
\ee
with the same notation for $+0$ as above.

This tells us that, according to the stability condition (5), only a
three-dimensional harmonic trap can house stable ideal Bose-condensed
gas. Bose-Einstein condensation cannot occur in one-dimensional
harmonic traps. And two-dimensional gas in a harmonic trap is weakly
unstable.

\subsection{Interacting nonuniform gas}

Let us consider an arbitrary nonuniform Bose system of atoms interacting
through repulsive forces. The general expression, characterizing particle
fluctuations, is given [4,17] by the variance
\be
\label{35}
 {\rm var}(\hat N) = N + \int \rho(\br) \rho(\br')
[ g(\br,\br') - 1 ] \; d\br d\br' \; ,
\ee
where the total particle density
\be
\label{36}
\rho(\br) = \rho_0(\br) + \rho_1(\br)
\ee
is the sum of the condensate density $\rho_0(\bf r)$ and the density
of uncondensed particles $\rho_1(\bf r)$, and $g(\bf r, \bf r')$ is
the pair correlation function. This expression is valid for any system
whether equilibrium or not.

For an equilibrium system, we shall use the local-density approximation
[17] in the frame of the self-consistent mean-field approach [18-22].
Then the density of uncondensed particles can be written [17] in the
form
\be
\label{37}
\rho_1(\br) = \int n(\bk,\br) \; \frac{d\bk}{(2\pi)^3} \; ,
\ee
in which
$$
 n(\bk,\br) = \frac{\om(\bk,\br)}{2\ep(\bk,\br)} \;
{\rm coth} \left [ \frac{\ep(\bk,\br)}{2T} \right ] \; - \;
\frac{1}{2}
$$
is the local momentum distribution, the notation
$$
\om(\bk,\br) \equiv mc^2(\br) + \frac{k^2}{2m}
$$
is used, and
$$
\ep(\bk,\br) =
\sqrt{ c^2(\br) k^2 + \left ( \frac{k^2}{2m} \right )^2}
$$
is the local Bogolubov spectrum. The local sound velocity is defined
by the
equation
\be
\label{38}
 mc^2(\br) = [\rho_0(\br) + \sgm_1(\br) ] \Phi_0 \; ,
\ee
where the interaction strength is given by
$$
 \Phi_0 \equiv \int \Phi(\br) \; d\br =
4 \pi \; \frac{a_s}{m} > 0 \; ,
$$
with positive scattering length $a_s$. And the anomalous average
\be
\label{39}
\sgm_1(\br) = \int \sgm(\bk,\br) \; \frac{d\bk}{(2\pi)^3}
\ee
is expressed through
$$
\sgm(\bk,\br) = - \;\frac{mc^2(\br)}{2\ep(\bk,\br)}
{\rm coth} \left [ \frac{\ep(\bk,\br)}{2T} \right ] \; .
$$
Substituting in the right-hand side of variance (35) the Bogolubov
shift [15] and keeping the terms up to second order with respect to
the operators of uncondensed particles [4,17], we have
$$
{\rm var}(\hat N) = N + 2 \lim_{k\ra 0} \int \rho(\br)
[ n(\bk,\br) + \sgm(\bk,\br) ] \; d\br \; .
$$
Accomplishing here the limit $k \ra 0$ yields
\be
\label{40}
 {\rm var}(\hat N) = \frac{T}{m}
\int \frac{\rho(\br)}{c^2(\br)}\; d\br \; .
\ee

Bose-Einstein condensation is accompanied by the global gauge
symmetry breaking, the latter being the necessary and sufficient
condition for the former [8,23]. When Bose-Einstein condensate is
present, the sound velocity $c(\bf r)$ is nonzero. Then the integral
in Eq. (40) is proportional to $N$. Therefore the fluctuation index,
describing particle fluctuations, is $\phi(\hat{N}) = 1$. The latter
means that for an arbitrary nonuniform three-dimensional system of
repulsive atoms, with Bose-Einstein condensate, particle fluctuations
are thermodynamically normal.

\section{Conclusion}

The notion of fluctuation indices for operators, representing
observable quantities, is introduced, characterizing the
stability properties of statistical systems. Thermodynamically
anomalous fluctuations imply the system instability, while stable
systems exhibit thermodynamically normal fluctuations. The notion
is illustrated by calculating the fluctuation indices for the
number-of-particle operator for systems with Bose-Einstein
condensate. It is shown that the ideal uniform Bose-condensed gas
is unstable. The ideal gas can be stabilized in trapping potentials,
provided that the confining dimension is larger than two. The case
of the confining dimension two is marginal, corresponding to a weakly
unstable system. Atomic gases in harmonic traps are stable only in the
three-dimensional case. Trapped one-dimensional gases are unstable, and
trapped two-dimensional gases are weakly unstable. A three-dimensional
Bose-condensed system of atoms, interacting through repulsive forces,
is stable for arbitrary external potentials.

The particular nature of a condensate can be different. For short,
we have been talking of atomic condensates. But there exist now several
types of molecular condensates (see Refs. [24-28] and review [10]).
The above consideration can be applied for any type of Bose condensates,
whether atomic or molecular.

Only equilibrium systems have been considered in the paper.
Fluctuations in nonequilibrium systems is a different topic
(see, e.g., [29-35]). But the notion of fluctuation indices can be
applied to nonequilibrium systems as well, since the definition of
fluctuation indices (4) is equally valid for any type of averages,
whether over equilibrium or nonequilibrium ensembles. The principal
difference between equilibrium and nonequilibrium systems is that
the stability conditions (5), generally, are not required for the
latter. The strength of fluctuations in nonequilibrium systems can
be of arbitrary magnitude. Though some restrictions on the fluctuation
strength could be connected with steady and quasiequilibrium states.

The fluctuation indices, introduced in the present paper, provide a
quantitative characteristic of fluctuation strength, associated with
the operators of observables. The knowledge of these indices can help
for deciding under what conditions and for what kind of traps one could
realize stable atomic systems in experiment.

\vskip 5mm

{\it Acknowledgement}

Financial support from the Russian Foundation for Basic Research is
acknowledged.

\end{document}